\documentclass{emulateapj}

\slugcomment{Accepted for publication in the Astrophysical Journal}

\shorttitle{Amplitude of solar p-modes from 3D simulations}
\shortauthors{Zhou, Asplund and Collet}

\usepackage{amsmath,amssymb}
\usepackage[normalem]{ulem}
\usepackage{bm}
\usepackage{comment}
\usepackage{xcolor}
\usepackage{soul}
\usepackage{hyperref}
\usepackage[percent]{overpic}
\usepackage{overpic}
\usepackage{subfigure}
\usepackage{natbib}

\begin{document}

\newcommand{\bea}{\begin{eqnarray}}
\newcommand{\eea}{\end{eqnarray}}
\newcommand{\E}{\mathrm{E}}
\newcommand{\Var}{\mathrm{Var}}
\newcommand{\bra}[1]{\langle #1|}
\newcommand{\ket}[1]{|#1\rangle}
\newcommand{\braket}[2]{\langle #1|#2 \rangle}
\newcommand{\mean}[2]{\langle #1 #2 \rangle}
\newcommand{\be}{\begin{equation}}
\newcommand{\ee}{\end{equation}}	
\newcommand{\ba}{\begin{eqnarray}}
\newcommand{\ea}{\end{eqnarray}}
\newcommand{\SD}[1]{{\color{magenta}#1}}
\newcommand{\rem}[1]{{\sout{#1}}}
\newcommand{\alert}[1]{\textbf{\color{red} \uwave{#1}}}
\newcommand{\Y}[1]{\textcolor{blue}{#1}}
\newcommand{\R}[1]{\textcolor{red}{#1}}
\newcommand{\B}[1]{\textcolor{black}{#1}}
\newcommand{\C}[1]{\textcolor{cyan}{#1}}
\newcommand{\db}{\color{darkblue}}

\newcommand{\ques}[1]{\textcolor{red}{#1}}
\newcommand{\yz}[1]{\textcolor{cyan}{#1}}
\newcommand{\ma}[1]{\textcolor{green}{#1}}
\newcommand{\rc}[1]{\textcolor{blue}{#1}}
\newcommand{\ds}[1]{\textcolor{orange}{#1}}

\newcommand{\ac}[1]{\textcolor{cyan}{\sout{#1}}}
\newcommand{\intinfty}{\int_{-\infty}^{\infty}\!}
\newcommand{\Tr}{\mathop{\rm Tr}\nolimits}
\newcommand{\const}{\mathop{\rm const}\nolimits}

\title{The amplitude of solar p-mode oscillations from three-dimensional convection simulations}

\author{Yixiao Zhou  \altaffilmark{1},
        Martin Asplund  \altaffilmark{1},
        and Remo Collet \altaffilmark{2}
}

\altaffiltext{1}{Research School of Astronomy and Astrophysics, Australian National University, Canberra, ACT 2611, Australia}
\altaffiltext{2}{Stellar Astrophysics Centre, Department of Physics and Astronomy, Ny Munkegade 120, Aarhus University, DK-8000 Aarhus C, Denmark}

\begin{abstract}

  The amplitude of solar p-mode oscillations is governed by stochastic excitation and mode damping, both of which take place in the surface convection zone. However, the time-dependent, turbulent nature of convection makes it difficult to self-consistently study excitation and damping processes through the use of traditional one-dimensional hydrostatic models. To this end, we carried out \textit{ab initio} three-dimensional, hydrodynamical numerical simulations of the solar atmosphere to investigate how p-modes are driven and dissipated in the Sun. The description of surface convection in the simulations is free from the tuneable parameters typically adopted in traditional one-dimensional models. Mode excitation and damping rates are computed based on analytical expressions whose ingredients are evaluated directly from the three-dimensional model. With excitation and damping rates both available, we estimate the theoretical oscillation amplitude and frequency of maximum power, $\nu_{\max}$, for the Sun. We compare our numerical results with helioseismic observations, finding encouraging agreement between the two. The numerical method presented here provides a novel way to investigate the physical processes responsible for mode driving and damping, and should be valid for all solar-type oscillating stars.
 
\end{abstract}


\section{Introduction} \label{sec:intro}

  Asteroseismology, the study of stellar oscillations, opens a unique window to probe various properties of stars. The analysis of the power spectra of stellar oscillations makes allows to infer the structure of stellar interiors as well as the fundamental parameters of stars \citep{2009ApJ...699.1403B,1995A&A...293...87K}. In some cases, it is possible to determine the evolutionary stages \citep{2012A&A...540A.143M}, size of surface/core convection zone \citep{1997MNRAS.287..189B,2016A&A...589A..93D}, or rotation rates \citep{2012Natur.481...55B} of stars from their oscillation frequencies. Among many applications, the asteroseismic determination of fundamental stellar parameters such as masses and radii is of great significance not only to stellar physics but to astronomy in general. For instance, stellar radii are used to derive exoplanet radii from the analysis of exoplanet transit light curves. Stellar masses can be related to stellar ages, which play a fundamental role in Galactic archaeology.
  
  Over the past decade, oscillations in thousands of solar-type stars have been detected by the CoRoT \citep{2008Sci...322..558M} and \textit{Kepler} \citep{2010Sci...327..977B} satellites, their number being destined to grow thanks to the current TESS \citep{2015JATIS...1a4003R} mission. These space-borne telescopes have paved the way to a new epoch of ensemble asteroseismology. In this context, the asteroseismic determination of stellar parameters from special empirical seismic scaling relations \citep{1991ApJ...368..599B,1995A&A...293...87K} has proven to be very effective. The seismic scaling relations link key seismic observables --large frequency separation $\Delta\nu$ and frequency of maximum oscillation power $\nu_{\max}$-- to stellar mass, radius and effective temperature, relatively to the Sun.
  
  The widespread application of the seismic scaling relations to stellar parameter determinations calls for a deeper investigation of the underlying physical processes behind the emergence of $\Delta\nu$ and $\nu_{\max}$.
  On the one hand,  $\Delta\nu$, a comparatively well understood quantity, is a proxy for the mean density of star \citep{1986ApJ...306L..37U}. On the other hand, $\nu_{\max}$, which is determined from oscillation amplitudes, has a tight connection with the driving and damping mechanisms of oscillations but its exact origin is still poorly understood. Although noticeable progress has been made towards the theoretical understanding of $\nu_{\max}$ \citep{2011A&A...530A.142B}, a complete solution to this issue still requires a thorough explanation of how modes are excited and damped in solar-type oscillators.
  
  Previous investigations have provided insights to the physics of both mode excitation and damping. \cite{1977ApJ...212..243G} first proposed that oscillations in the Sun are driven by turbulent convection. From this promise, \cite{1994ApJ...424..466G} and \cite{2001A&A...370..136S} developed theoretical formulations to model the stochastic excitation of oscillations by stellar surface convective motions, assuming a one-dimensional (1D) description of convection based on the mixing-length theory. Based on these theoretical prescriptions, \cite{1994ApJ...424..466G} and \cite{2010A&A...522L...2B} computed numerically the excitation rates for the Sun, finding a satisfactory match to observationally inferred values. In addition, \cite{2008A&A...489..291S} showed that the same theoretical framework for mode excitation is also applicable to other solar-like oscillators such as $\alpha$ Centauri A. On the other side, it is natural to analyse the stability of these stochastically-driven modes, and, if stable, the rate of their energy dissipation. The first question was examined in detail by \cite{1992MNRAS.255..603B} and the conclusion was that all solar p-modes are stable. The second question was answered by \cite{1999A&A...351..582H} and \cite{2005MNRAS.360..859C}, who computed damping rates for solar p-modes using a non-local mixing-length model, finding good agreement with observed line widths (see \citealt{2015LRSP...12....8H} for a review of the model). 
  
  There is no doubt that the aforementioned studies have greatly enriched our understanding toward the nature of p-mode oscillations in solar-type stars. However, due to the lack of highly realistic analytical description of convection and turbulence, theoretical models adopted in these works are unavoidably simplified to some extent. In particular, the dynamic and chaotic nature of turbulence is embedded in free parameters that have to be calibrated either from other theoretical models or from observational data. A promising approach to overcome this difficulty is to simulate excitation and damping of modes from first principle using three-dimensional (3D) hydrodynamical numerical simulations. Noticeable breakthrough in this direction has already been made by \cite{2001ApJ...546..576N} and \cite{2001ApJ...546..585S}. In their pioneering work, the excitation rates of solar radial modes are extracted directly from 3D simulation of near-surface layers of the solar convective region. Their results involve no free parameters and fall in line with observation. In this paper, we expand their idea and further explore the possibility of modelling both mode excitation (Sect.~\ref{sec:excitation}) and damping (Sect.~\ref{sec:damping}) using 3D simulation. Knowledge of how modes are driven and dissipated enables us to estimate their amplitude, from which a theoretical $\nu_{\max}$ value can be deduced (Sect.~\ref{sec:Vandnumax}). The results of these \textit{ab initio} parameter-free numerical calculations are compared with helioseismic observations. As a first attempt on this topic, we focus on the Sun and limit our discussion to radial modes only.

\section{Three dimensional solar atmosphere model} \label{sec:model}

\begin{figure}
\begin{overpic}[width=0.95\columnwidth]{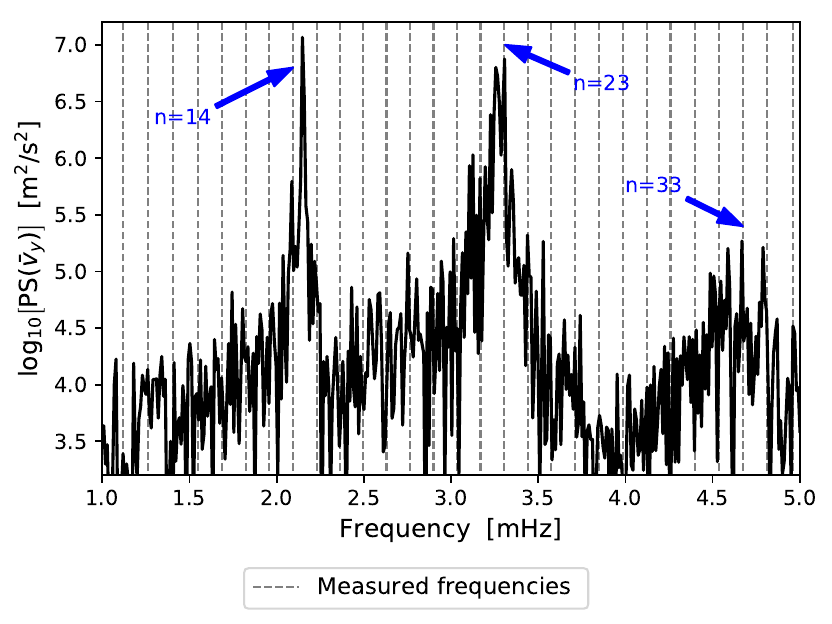}
\end{overpic}
\caption{Vertical velocity power spectrum computed from 3D solar simulation, from which three simulation modes with cyclic frequency 2.148, 3.307 and 4.668 mHz are recognisable. Measured solar radial p-mode frequencies (degree $l=0$, vertical grey dashed lines) are shown in the background with $n$ denotes mode radial order. The observed frequency values are from Global Oscillations at Low Frequency (GOLF, \citealt{2001SoPh..200..361G,2002A&A...394..285G}).}
\label{fig:PSvy}
\end{figure}

  In this section, we briefly describe the 3D hydrodynamical model solar atmosphere used in this work. The model is computed with the \textsc{Stagger} code \citep{1995...Staggercodepaper,2018MNRAS.475.3369C}, a state-of-the-art, radiative-magnetohydrodynamics code that solves the equations of mass, momentum, and energy conservation, and magnetic-field induction equation in 3D. Radiative energy transport is modelled by solving the 3D equation of radiative transfer at every time-step of the simulation along different inclined rays in space. For the present solar simulation, nine directions are used, including the vertical and eight inclined directions (a combination of two polar $\theta$-angles and four azimuthal $\phi$-angles). The code also adopts realistic micro-physics. It uses a modified version of the \cite{1988ApJ...331..815M} equation of state \citep{2013ApJ...769...18T} that accounts for the 17 most abundant elements in the Sun as well as the $\rm H_2$ and $\rm H_2^+$ molecules (cf.~\citealt{2013ApJ...769...18T} Sect.~2.1). A comprehensive collection of relevant continuous absorption and scattering are included \citep{2010A&A...517A..49H}. Line opacities are treated using opacity binning \citep{1982A&A...107....1N,2013A&A...557A..26M}, with 12 opacity bins divided in both wavelength and strength of opacity.
  
  Our \textsc{Stagger} model stellar atmosphere simulates a small part of the Sun near photosphere. It assumes a constant gravitational acceleration and ignores magnetic field. The simulation domain covers 6 Mm $\times$ 6 Mm area horizontally, and 3.8 Mm in the vertical (radial) direction -- approximately $1$ Mm above the base of the photosphere and $2.8$ Mm below it. The domain is discretized on a 3D Cartesian grid, with spatial resolution $240^3$, which is sufficient for studying mode excitation\footnote{As discussed in \cite{2007A&A...463..297S}, the differences between excitation rates computed based on $253 \times 253 \times 163$ and  $125 \times 125 \times 82$ solar simulations are small, implying our adopted spatial resolution is sufficient for this problem.}.
Scalars, such as densities and internal energies, are evaluated at cell center while vectors, such as momentum densities, are staggered at cell faces (hence the meaning of ``Stagger''). The simulation domain is small but nevertheless representative, because it resides in the region where 3D effects, such as fluctuations in horizontal plane caused by up and down flow of fluid and the presence of strong turbulence, are most prominent. We note also that spherical effect in our model is negligible because the vertical scale is small compared to the total radius of the Sun. Temporally, our model spans 24 hours solar time, with one snapshot stored every 30 seconds. This sampling interval is adequately short, since doubling it will not influence our excitation and damping results (see in Sect.~\ref{sec:excitation} and \ref{sec:damping}) significantly.
  
  Global parameters of our solar model are very close to actual solar values. The mean effective temperature over 24 hours solar time is $5772.7$ K, almost the same as the nominal solar value ($5772.0$ K, \citealt{2016AJ....152...41P}); gravitational acceleration is set to $2.74 \times 10^4 \; \rm g/cm^2$, taken from \cite{2016AJ....152...41P}; for element abundances we adopt the \cite{2009ARA&A..47..481A} solar composition.
  
  P-mode oscillations are natural phenomena in our model. Radial p-modes can be identified by looking at the power spectrum of horizontally averaged vertical velocity:
\begin{equation}
{\rm PS}[\bar{v}_y](\omega) = \frac{1}{N}
\left\vert \sum\limits_{s=0}^{N-1} \bar{v}_{y}(t_s) e^{i\omega s \Delta\tau} \right\vert^2.
\end{equation}
Here $\bar{v}_{y}$ is the horizontally averaged vertical velocity of snapshot $s$, $\omega$ is angular frequency, $\Delta\tau$ being the time interval between two consecutive snapshots, $N$ the total snapshot number ($N=2880$ in our case). The power spectrum of $\bar{v}_y$ around photosphere is shown in Fig.~\ref{fig:PSvy}, the peaks represent radial p-mode naturally emerged in the simulation box. In total three simulation modes with frequency 2.148, 3.307, 4.668 mHz are identifiable from Fig.~\ref{fig:PSvy}; no obvious oscillation signature is found below 1.5 mHz or above 5 mHz. Two of them have frequencies that are close to the p-mode frequencies  measured in the Sun. The simulation mode with lowest frequency deviates from the observed quantity. This reflects the finite extent of the 3D simulation, since lower frequency modes have greater amplitudes at depth than higher frequency ones as seen in Fig.~\ref{fig:xi}. Had the 3D model extended to deeper interior of the Sun, the expectation is that this discrepancy would be reduced.
  
  Although the property of simulation modes are affected by the finite spatial dimension of the simulation box\footnote{The property of simulation modes, as well as $\bar{v}_y$, will not enter into the subsequent computation of excitation rates, because the main effect of p-mode oscillations in the simulation is filtered out by mapping physical quantities into pseudo-Lagrangian frame (cf.~Sect.~\ref{sec:er_num}).} (hence also called ``box modes''), we clarify that ``box modes'' are natural phenomena in 3D simulation rather than numerical noise. We further clarify the peaks in Fig.~\ref{fig:PSvy} represent radial modes rather than the radial component of low-degree non-radial modes, although their oscillation frequencies can be close to each other. The degree $l$ of a mode indicates the number of its surface nodes (fixed point from the north pole, along the surface of the sphere, to south pole). The horizontal wavelength $\lambda_h$ and the degree of a mode are related by:
\begin{equation} \label{eq:wlandl}
\frac{\lambda_h}{2} \simeq \frac{\pi\mathcal{R}}{l},
\end{equation}
where $\mathcal{R}$ is the stellar radius. In order to resolve a non-radial mode in the simulation, the horizontal wavelength of this mode should be shorter than the horizontal domain of the simulation. Therefore, in our case, non-radial oscillations can be identified only if its horizontal wavelength $\lambda_h \lesssim 6$ Mm, which corresponds to $l \gtrsim 729$. In other words, non-radial low-degree (i.e.~$l = 1,2,3$ ...) p-modes cannot be detected because of the limited simulation domain.

  \cite{1998ApJ...499..914S} have demonstrated that the granulation pattern from 3D simulation of solar atmosphere strongly resembles what observed on the Sun, after considering telescope and atmosphere seeing. The distribution of granule size from simulations is in agreement with solar observation as well \citep{1998ApJ...499..914S}. \cite{2000A&A...359..729A} and \cite{2013A&A...554A.118P} also reported excellent match when comparing the detailed spectral lines predicted from 3D solar simulation with observation.

  These facts lead us to believe that such 3D solar atmosphere model is a highly realistic representation of the physical processes taking place near the solar surface region. Our 3D \textsc{Stagger} solar model will be applied to the subsequent calculation of excitation rates.

\section{Excitation rates} \label{sec:excitation}

  In the Sun, the observed p-mode oscillations are driven by near-surface convection. The driving process is quantified by (mode) excitation rate which describes how fast energy is supplied from stochastic convection to coherent fluid motion. In this section, we will specify how to extract excitation rates from 3D stellar atmosphere model (Sect.~\ref{sec:er_theory}, \ref{sec:er_num}), and present numerical results for the Sun (Sect.~\ref{sec:er_result}). Here we confine the discussion to radial oscillations.

\subsection{Theoretical formulation} \label{sec:er_theory}

  The expression of excitation rate for radial modes was originally derived in pioneering works by \cite{1994ApJ...424..466G}, \cite{2001A&A...370..136S} and \citet[abbreviated NS01 hereinafter]{2001ApJ...546..576N}. We follow the formulation developed in NS01, since it is more suitable for direct numerical evaluation. NS01 started with basic fluid equations in 3D, rewrote them to horizontal-averaged perturbed fluid equations. From the (1D) perturbed equations they obtained the expression of work integral (defined below) with proper approximation, then arrived at the change in mode kinetic energy per unit area over certain time interval $\Delta t$ (i.e., excitation rate per unit area, NS01 Eqs.~65, 66 and 71)
\begin{equation} \label{eq:er_orig}
\begin{aligned}
\frac{\Delta \langle E_{\omega} \rangle_{\rm ens}}{\Delta t} 
= \frac{1}{4\Delta t} \left\langle \left\vert
\int_{\Delta t} \int_r \delta \bar{P}_{\rm nad}(r,t) 
\frac{1}{E_0^{1/2}}\frac{\partial (\dot{\delta r})}{\partial r} dr \: dt
\right\vert^2 \right\rangle_{\rm ens},
\end{aligned}
\end{equation}
with $E_0$ the mode kinetic energy per unit surface area (NS01 Eq.~63):
\begin{equation} \label{eq:E_0}
E_0 = \frac{\omega^2}{2} 
\int_r \rho \xi_r^2(r) \left(\frac{r}{R_{\rm phot}}\right)^2 \: dr.
\end{equation}
The integral over radius $r$ in Eq.~\eqref{eq:er_orig} is the so-called ``work integral'', $\rho$ is density, $R_{\rm phot}$ is photosphere radius, and $\delta \bar{P}_{\rm nad}$ is the horizontally averaged non-adiabatic pressure fluctuation that arises from non-adiabatic effects including entropy fluctuation and convective turbulence. The $\langle ... \rangle_{\rm ens}$ bracket stands for the ensemble average over all phases (NS01 Sect.~3.2), it is necessary because the phase differences between coherent waves and chaotic convective processes are completely random. The canonical form of mode displacement vector $\delta\vec{r}$ is written as (we refer the readers to \citealt{2010aste.book.....A} Sect.~3.3.1 for a thorough introduction)
\begin{equation} \label{eq:mode_disp}
\begin{aligned}
& \delta\vec{r} = \mathfrak{Re}\left\lbrace 
\left[ \xi_r(r) Y_{lm}(\theta,\phi) \vec{e}_r + \right.\right.
\\
& \left.\left. \xi_h(r) \left( \partial_\theta Y_{lm}(\theta,\phi) \vec{e}_{\theta} + 
\frac{1}{\sin\theta} \partial_\phi Y_{lm}(\theta,\phi) \vec{e}_{\phi} \right)
\right]e^{-i(\omega t + \varphi)} \right\rbrace,
\end{aligned}
\end{equation}
where $\xi_r$ and $\xi_h$ are the amplitude functions (also named mode eigenfunctions from numerical point of view), and $Y_{lm}(\theta,\phi)$ are the spherical harmonic functions with $l,m$ being the angular quantum numbers\footnote{$\mathfrak{Re}\lbrace f \rbrace$ ($\mathfrak{Im}\lbrace f \rbrace$) means the real (imagery) part of complex function $f$.}. Note that $\varphi$ is an arbitrary phase factor, however, in our context, it is the phase difference between non-adiabatic pressure fluctuation and oscillation mode. As we are considering radial modes only, Eq.~\eqref{eq:mode_disp} simplifies into
\begin{equation} \label{eq:mode_disp_simp}
\begin{aligned}
\delta r = \mathfrak{Re} \left\lbrace \xi_r(r) e^{-i(\omega t + \varphi)} \right\rbrace
\end{aligned}.
\end{equation}
In Eq.~\eqref{eq:er_orig}, the coupling between $\delta\bar{P}_{\rm nad}$ and $\delta r$ reflects the interaction between convection and pulsation which is responsible for mode excitation. Substituting Eq.~\eqref{eq:mode_disp_simp} into Eq.~\eqref{eq:er_orig},
\begin{equation} \label{eq:er_interm}
\begin{aligned}
\frac{\Delta \langle E_{\omega} \rangle_{\rm ens}}{\Delta t} 
&= \frac{1}{4\Delta t} \left\langle \left\vert \mathfrak{Re} \left\{
\int_r  (-i\omega) e^{-i\varphi} \frac{1}{E_0^{1/2}}\frac{\partial \xi_r}{\partial r}
\right.\right.\right. 
\\
&\left.\left.\left. 
\int_{\Delta t}  \delta\bar{P}_{\rm nad}(r,t) e^{-i\omega t} \; dt  \; dr 
\right\} \right\vert^2 \right\rangle_{\rm ens}.
\end{aligned}
\end{equation}
The time integral in Eq.~\eqref{eq:er_interm} is equivalent to the Fourier transform of non-adiabatic pressure fluctuation. Expanding Eq.~\eqref{eq:er_interm} gives
\begin{equation} \label{eq:er_expan}
\begin{aligned}
& \frac{\Delta \langle E_{\omega} \rangle_{\rm ens}}{\Delta t} =
\\
& \frac{\omega^2}{4 \Delta t} \left\langle
\sin^2\varphi \left(\int_r \frac{1}{E_0^{1/2}}\frac{\partial \xi_r}{\partial r} 
\mathfrak{Re}\left\lbrace \mathcal{F}[\delta \bar{P}_{\rm nad}] \right\rbrace \: dr \right)^2 \right\rangle_{\rm ens} 
\\
& - \frac{\omega^2}{4 \Delta t} \left\langle
2\sin\varphi\cos\varphi \left(\int_r \frac{1}{E_0^{1/2}}\frac{\partial \xi_r}{\partial r} 
\mathfrak{Re}\left\lbrace \mathcal{F}[\delta \bar{P}_{\rm nad}] \right\rbrace \: dr \right) \right.
\\
& \left.\left(\int_r \frac{1}{E_0^{1/2}}\frac{\partial \xi_r}{\partial r} 
\mathfrak{Im}\left\lbrace \mathcal{F}[\delta \bar{P}_{\rm nad}] \right\rbrace \: dr \right) \right\rangle_{\rm ens} 
\\
& + \frac{\omega^2}{4 \Delta t} \left\langle
\cos^2\varphi \left(\int_r \frac{1}{E_0^{1/2}}\frac{\partial \xi_r}{\partial r} 
\mathfrak{Im}\left\lbrace \mathcal{F}[\delta \bar{P}_{\rm nad}] \right\rbrace \: dr \right)^2 \right\rangle_{\rm ens}.
\end{aligned}
\end{equation}
The ensemble average over all phases is calculated as
\begin{equation}
\langle f \rangle_{\rm ens} = \frac{1}{2\pi}\int_0^{2\pi} f(\varphi) \: d\varphi.
\end{equation}
Therefore, $\langle \sin^2\varphi \rangle_{\rm ens} = 1/2$, $\langle \cos^2\varphi \rangle_{\rm ens} = 1/2$, $\langle 2\sin\varphi \cos\varphi \rangle_{\rm ens} = 0$ and Eq.~\eqref{eq:er_expan} simplifies into
\begin{equation} \label{eq:er_fi}
\begin{aligned}
\frac{\Delta \langle E_{\omega} \rangle_{\rm ens}}{\Delta t} 
=& \frac{\omega^2}{8 \Delta t} \left[
\left(\int_r \frac{1}{E_0^{1/2}} \frac{\partial \xi_r}{\partial r} \mathfrak{Re}\left\lbrace \mathcal{F}[\delta \bar{P}_{\rm nad}] \right\rbrace \: dr \right)^2  \right.
\\
+& \left. \left(\int_r \frac{1}{E_0^{1/2}} \frac{\partial \xi_r}{\partial r} 
\mathfrak{Im}\left\lbrace \mathcal{F}[\delta \bar{P}_{\rm nad}] \right\rbrace \: dr \right)^2 \right],
\end{aligned}
\end{equation}
where $\mathcal{F}$ represents Fourier transform from time to frequency domain. Eq.~\eqref{eq:er_fi} is essentially equivalent to Eq.~5 in \citet[SN01]{2001ApJ...546..585S}, we will use \eqref{eq:er_fi} to calculate excitation rates.

\subsection{Numerical evaluation} \label{sec:er_num}

\begin{figure}
\begin{overpic}[width=0.95\columnwidth]{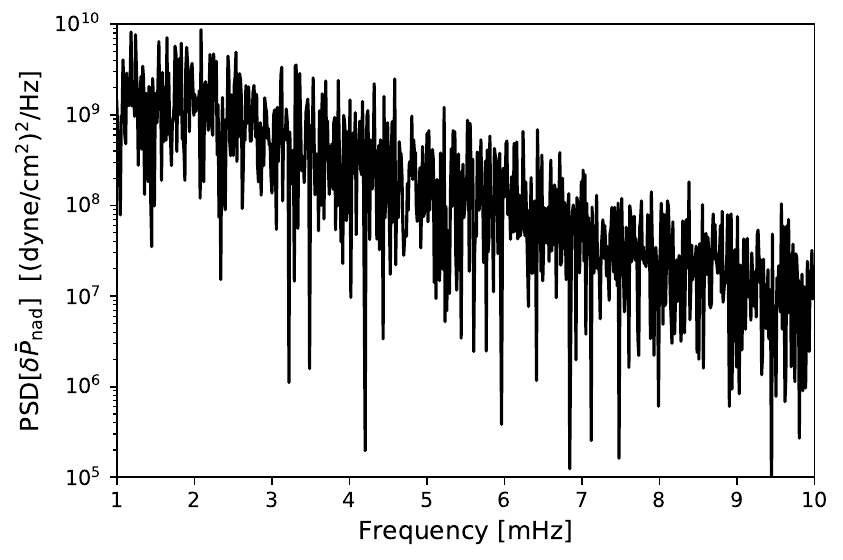}
\end{overpic}
\caption{$\delta\bar{P}_{\rm nad}$ power spectrum density at $100$ km below photosphere, as computed via Eq.~\eqref{eq:PSD_deltaP}. }
\label{fig:PSdeltaP}
\end{figure}

\begin{figure}
\begin{overpic}[width=0.95\columnwidth]{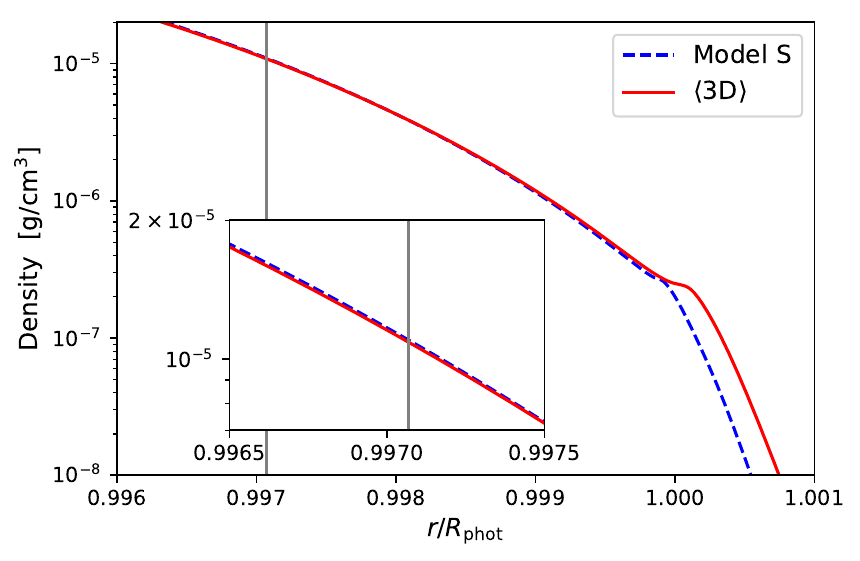}
\end{overpic}
\caption{The predicted density profile in near-surface region of solar models as a function of fractional radius (normalized by radius at photosphere), with a zoom-in near matching point. Blue dashed and red solid lines represent model S and horizontal- and time-averaged 3D model respectively. Grey vertical line marks the location of interior matching point.}
\label{fig:compare_rho}
\end{figure}

\begin{figure}
\begin{overpic}[width=0.95\columnwidth]{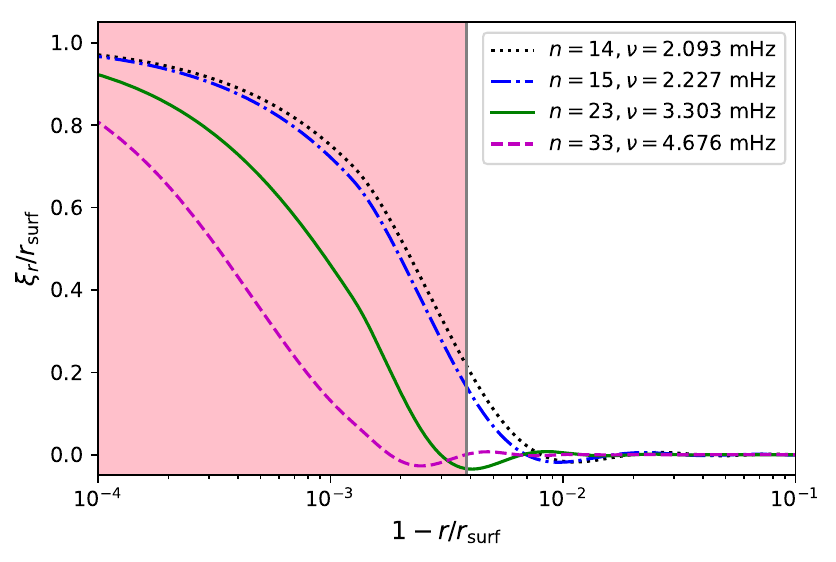}
\end{overpic}
\caption{Normalized eigenfunctions for four radial modes computed from \textsc{adipls} with the patched solar model. Here $r_{\rm surf}$ is the radius of the upper-most (surface) point of the patched model. Theoretical frequencies and radial orders of these example modes are also shown. Pink-shaded zone is the part covered by the 3D simulation. Note that the magnitude of eigenfunction in solar interior, relative to its surface value, decreases with increasing frequency.
}
\label{fig:xi}
\end{figure}

  Two key ingredients in Eq.~\eqref{eq:er_fi} are non-adiabatic pressure fluctuation $\delta\bar{P}_{\rm nad}$ and amplitude function $\xi_r$, representing the dynamics of convection and oscillation respectively. In this subsection, we explain how they are computed numerically.
  
  The time-dependent nature of 3D model allows the direct evaluation of $\delta\bar{P}_{\rm nad}$, which is the difference between total and adiabatic pressure fluctuation (SN01 Eq.~3),
\begin{equation} \label{eq:SN01eq3}
\begin{aligned}
\delta\bar{P}_{\rm nad} = \delta\bar{P} - \delta\bar{P}_{\rm ad}
= (\delta\ln \bar{P} - \bar{\Gamma}_1 \delta\ln \bar{\rho}) \bar{P},
\end{aligned}
\end{equation}  
where $\delta$ stands for Lagrangian perturbation. $P$ and $\Gamma_1$ are total pressure and first adiabatic index respectively. As before, the bar symbol denotes horizontal averaging. In 3D hydrodynamical stellar atmosphere simulations, the change of physical quantities around their mean value consists of mainly two contributions: (a) perturbations from radiative and convective processes (b) collective fluid motions in the simulation domain (i.e.~p-mode oscillations). Because (b) is approximately adiabatic, it is important to isolate (a) from (b) in order to calculate $\delta\bar{P}_{\rm nad}$. This is achieved by mapping variables from the original Cartesian frame to a frame that is co-moving with collective fluid motion. The latter is often named pseudo-Lagrangian frame, which is characterized by horizontal- and time-averaged column mass density at each geometric depth,
\begin{equation}
\sigma(y) = \left\langle 
\int_{y_{\rm top}}^{y} \bar{\rho}(y^\prime,t_s) \, dy^\prime \right\rangle_t,
\end{equation}
which filters out the main effect of radial p-modes \citep{1999A&A...351..689R,2014MNRAS.445.4366T}. Here, $\langle ... \rangle_t$ means time average (that is, average over all snapshots), $y_{\rm top}$ is the geometric depth at the top of simulation domain, and the $s$ index refers to the snapshot number. In pseudo-Lagrangian frame, the Lagrangian perturbation changes into Eulerian perturbation\footnote{Detailed explanations to Eulerian and Lagrangian perturbations is available in, e.g., \cite{2010aste.book.....A} Chapter 3.}, hence Eq.~\eqref{eq:SN01eq3} can be simplified into
\begin{equation} \label{eq:deltaP_num}
\begin{aligned}
\delta \bar{P}_{\rm nad}(t) =&
\left[ \left( \ln\bar{P}_{\rm L}(t) - \ln\bar{P}_{0,\rm L} \right) \right.
\\ 
&- \left.\bar{\Gamma}_{1,\rm L} \left( \ln\bar{\rho}_{\rm L}(t) - \ln\bar{\rho}_{0,\rm L} \right) \right] \bar{P}_{\rm L}.
\end{aligned}
\end{equation}
Here quantities defined in pseudo-Lagrangian frame are marked with subscript ``L'', while the subscript ``0'' stands for the equilibrium state. Numerically, the equilibrium state is calculated by taking the temporal average over all simulation snapshots.

  Non-adiabatic pressure fluctuation is computed via \eqref{eq:deltaP_num}, its power spectrum density,
\begin{equation} \label{eq:PSD_deltaP}
{\rm PSD}[\delta\bar{P}_{\rm nad}](\omega) = \frac{\Delta\tau}{N} \left\vert \sum\limits_{s=0}^{N-1} \delta\bar{P}_{\rm nad}(t_s) e^{i\omega s \Delta\tau} \right\vert^2,
\end{equation}  
is depicted in Fig.~\ref{fig:PSdeltaP}. No obvious peak is observed in the figure, which suggests non-adiabatic pressure fluctuation is associated with turbulent convection, with no preference over any specific frequency. This is in accordance with the conclusion in SN01.
  
  \
  
  We now turn to the other component of Eq.~\eqref{eq:er_fi}, $\xi_r$, a quantity that is solely relevant to oscillation. In the case of radial mode, $\xi_r$ describes mode amplitude distribution in the star. Contrary to $\delta \bar{P}_{\rm nad}$, we choose to compute $\xi_r$ with patched 1D model rather than extract it from simulation. The reason is that, first of all, only three modes are clearly identifiable in 3D model, much less than the number of radial modes detected for the Sun. Secondly, although the oscillation amplitude of radial modes are largest in the near-surface region covered by the simulation, they actually propagate throughout the entire star. The patched 1D model is obtained by combining 1D interior model with horizontal- and time-averaged 3D model. Therefore, it extends from stellar center to the upper boundary of 3D model. Another advantage of patched model is that it reduces the ``surface effect''\footnote{The ``surface effect'' refers to the mismatch between observed and predicted p-mode oscillation frequencies at high frequencies in the Sun and other solar-type stars, which is due to the inadequate description of the outer stellar convection zone and atmospheric structure by traditional 1D models as well as the adiabatic assumption when computing theoretical mode frequencies. For efforts on this topic, consult, e.g., \cite{1999A&A...351..689R}, \cite{2017MNRAS.466L..43T}, \cite{2017MNRAS.472.3264J} and \cite{2017MNRAS.464L.124H}.}, thereby bringing theoretical p-mode frequencies closer to measured values.
  
  In this work, we adopt the standard solar model (also called model S, \citealt{1996Sci...272.1286C}) as the 1D interior model. Model S is widely used as a reference model for helioseismic analysis. The model-predicted sound speed and density profile are both in good agreement with corresponding heliosismic-inferred values (cf.~\citealt{1997MNRAS.292..243B} Figs.~6 and 10).
  
  To combine horizontal- and time-averaged 3D atmosphere model with model S, we first select a matching point in atmosphere model which is located slightly above the bottom of simulation domain (although not exactly \emph{at} the bottom layer, so to avoid artificial boundary effects) for the purposes of minimizing horizontal fluctuations in physical quantities. The matching point in model S is then determined by requiring the pressure to be identical to the average one at the matching point in the simulation, that is
\begin{equation}
\langle \bar{P}_{\rm 3D}(y_{\rm am}) \rangle_t = P_{\rm 1D}(r_{\rm im}),
\end{equation}
where $y_{\rm am}$ is the geometric depth of matching point in 3D model (``am'' stands for atmosphere matching point) and $r_{\rm im}$ is the radius of matching point in 1D model (``im'' stands for interior matching point). The matching procedure ensures a continuous transition in pressure between averaged 3D and 1D model, and provides a unified depth scale (radius) between the two. We caution that due to 3D effects and different micro-physics between 3D and 1D model, there might be discontinuities for other quantities, for instance density, at the matching point. Nevertheless, the discontinuities are found to be small (Fig.~\ref{fig:compare_rho}) hence not likely to affect our results significantly. Finally, we trim the atmosphere and interior model by discarding all points below the atmosphere matching point in the averaged 3D model, and all points above the interior matching point in 1D model, then put them together to get the patched solar model.

  We use the Aarhus adiabatic oscillation package (\textsc{adipls}, \citealt{2008Ap&SS.316..113C}) to compute theoretical p-mode frequencies and eigenfunctions, as well as mode kinetic energy $E_0$, with patched solar model as input. Numerical results are depicted in Fig.~\ref{fig:xi} for four example p-modes. Our main focuses are high order ($n>10$) radial ($l=0$) p-modes with cyclic frequency below the acoustic cut-off frequency ($\sim 5$ mHz for the Sun). Here we emphasize that eigenfunctions obtained from \textsc{adipls} are normalized by their surface value, thus their absolute value has no physical meaning. To facilitate comparison between eigenfunctions, we eliminate the dependence on the normalization condition by dividing by their kinetic energy $\sqrt{E_0}$. As an example, the squared normalized eigenfunction gradients, $(\partial_r\xi_r)^2/E_0$, of two example radial modes are demonstrated in Fig.~\ref{fig:Pandxi}, together with non-adiabatic pressure fluctuations at the same frequency.

\subsection{Results} \label{sec:er_result}

\begin{figure}
\begin{overpic}[width=0.95\columnwidth]{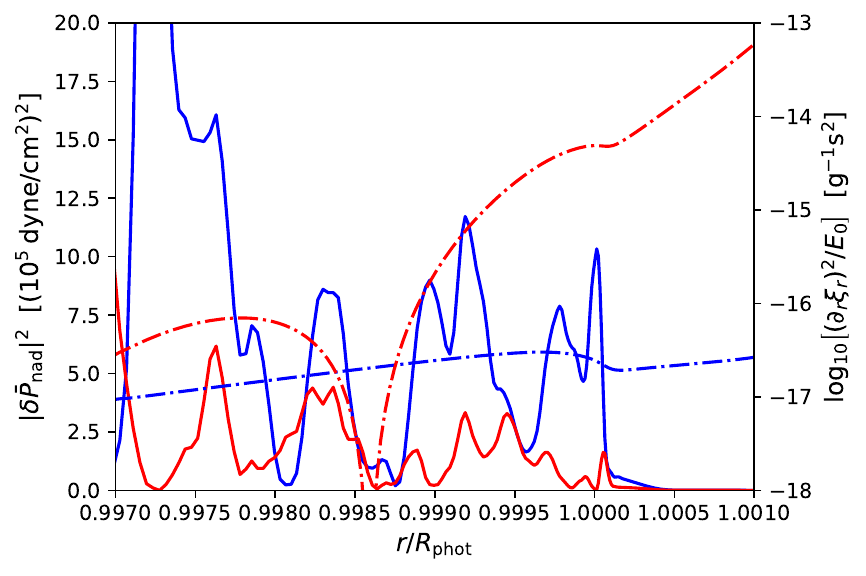}
\end{overpic}
\caption{Normalized eigenfunction gradient (dash-dot lines) in the outer part of solar model, as well as non-adiabatic pressure fluctuation distribution (solid lines) at same frequency. Blue color represents $n=11$, $\nu \approx 1.69$ mHz (lower frequency) radial mode while red corresponds to $n=30$, $\nu \approx 4.26$ mHz (higher frequency) radial mode. Radial orders and mode frequencies are determined from \textsc{adipls}.}
\label{fig:Pandxi}
\end{figure}

\begin{figure}
\begin{overpic}[width=0.95\columnwidth]{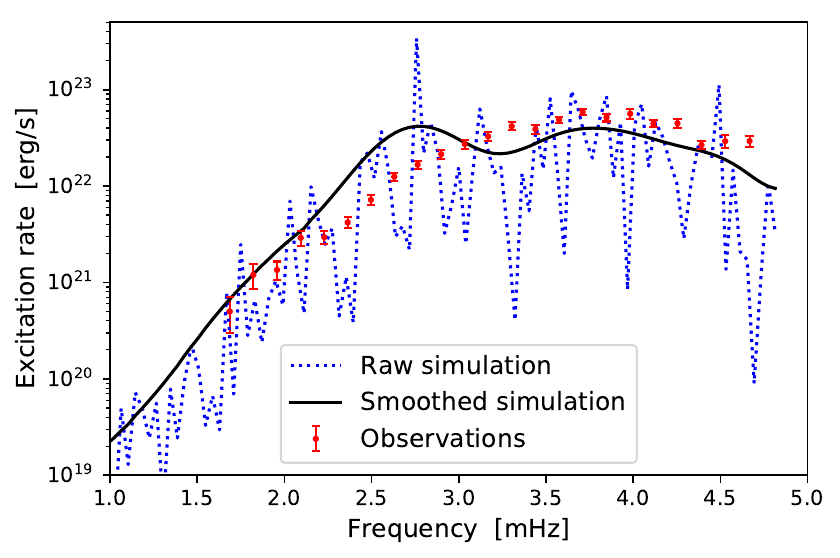}
\end{overpic}
\caption{Excitation rates as a function of cyclic frequency computed from 3D solar atmosphere model and 1D patched solar model. Original data from simulation are shown in blue dotted line, and black solid curve is the result after smoothing with Gaussian kernel with full width at half maximum equals to 0.38 mHz. Excitation rates for $l=0$ modes from Birmingham Solar-Oscillations Network (BiSON, \citealt{1998MNRAS.298L...7C}) are shown with red circles with uncertainties. The BiSON excitation rates have been divided by 2 in order to account for different definitions of mode energy (cf.~Eq.~\eqref{eq:def_Eosc} and \citealt{1998MNRAS.298L...7C}, Sect.~2.2).}
\label{fig:er}
\end{figure}

  Now that two key quantities appearing in Eq.~\eqref{eq:er_fi} have been computed, we proceed with the evaluation of the excitation rates. As $\delta\bar{P}_{\rm nad}$ is accessible in practice only through the 3D simulation, the lower integration limit in Eq.~\eqref{eq:er_fi} is the corresponding radius at the bottom of simulation domain, $r_{\rm 3D \: bot}$. On the other side, the work integral is terminated at the upper-most point $r_{\rm surf}$ of the patched model. Therefore Eq.~\eqref{eq:er_fi} finally becomes
\begin{equation} \label{eq:er_num}
\begin{aligned}
\frac{\Delta \langle E_{\omega} \rangle_{\rm ens}}{\Delta t} 
=& \frac{\omega^2}{8 \Delta t} \left[
\left(\int_{r_{\rm 3D \: bot}}^{r_{\rm surf}} \frac{1}{E_0^{1/2}}\frac{\partial \xi_r}{\partial r} 
\mathfrak{Re}\left\lbrace \mathcal{F}[\delta \bar{P}_{\rm nad}] \right\rbrace \: dr \right)^2  \right.
\\
+& \left. \left(\int_{r_{\rm 3D \: bot}}^{r_{\rm surf}} \frac{1}{E_0^{1/2}} \frac{\partial \xi_r}{\partial r} 
\mathfrak{Im}\left\lbrace \mathcal{F}[\delta \bar{P}_{\rm nad}] \right\rbrace \: dr \right)^2 \right],
\end{aligned}
\end{equation}
with kinetic energy integrated from stellar center to surface
\begin{equation}
E_0 = \frac{\omega^2}{2} 
\int_0^{r_{\rm surf}} \rho \xi_r^2(r) \left(\frac{r}{R_{\rm phot}}\right)^2 \: dr.
\end{equation}
Eq.~\eqref{eq:er_num} is evaluated at different angular frequencies. For frequency values not equal to mode eigenfrequencies, the eigenfunction is not directly available, so we linearly interpolate $\partial_r\xi_r$ to our target frequency value (see also the Appendix in SN01). Recall that $\Delta \langle E_{\omega} \rangle_{\rm ens} / \Delta t$ is the excitation rate per unit area, to get global excitation rate we multiply $\Delta \langle E_{\omega} \rangle_{\rm ens} / \Delta t$ by the horizontal area\footnote{Not by the total surface area of the Sun $4\pi R_{\rm phot}^2$; the underlying reason is explained in NS01 Sect.~3.3 and \cite{2004SoPh..220..229S} Sect.~5.} of the simulation, that is, $36$~$\rm Mm^2$.

  Global excitation rates from the simulation (both original and smoothed data) are displayed in Fig.~\ref{fig:er}, together with excitation rates extracted from observations \citep{1998MNRAS.298L...7C}. Broadly speaking, excitation rates predicted from simulation agree well with observations. The simulation result demonstrates a strong fluctuation with frequency, which stems from non-adiabatic pressure fluctuation (Fig.~\ref{fig:PSdeltaP}). Its overall trend is also clear from the smoothed curve --- excitation rate is small at lower frequencies, increases rapidly with increasing frequency, reaches a plateau that ranges from $\sim 2.5$ mHz to $\sim 4$ mHz, then starts to decline at higher frequencies. The trend observed in Fig.~\ref{fig:er} can be explained by analysing the behaviour of two key ingredients in Eq.~\eqref{eq:er_num}, namely $(\partial_r\xi_r)^2/E_0$ and $\delta\bar{P}_{\rm nad}$.
  
  The normalized eigenfunction gradient $(\partial_r \xi_r)^2/E_0$ quantifies the fluid's compression associated with oscillations: the larger the value of the gradient is, the stronger the compression locally. An extreme case is $\partial_r \xi_r = 0$, where the oscillation amplitude does not vary locally to first order and fluid elements move in sync and no compression occurs anywhere. As shown in Fig.~\ref{fig:Pandxi}, for the lower frequency mode, compression is weak throughout the simulation domain while for the higher frequency one, strong compression occurs around and above photosphere. 
  The reason for this is, as mentioned in Sect.~\ref{sec:model} and demonstrated in Fig.~\ref{fig:xi}, lower frequency p-modes have greater amplitude in the deep interior of star where density is significantly higher. As a result, if two modes have same kinetic energy, the lower frequency mode will show smaller amplitude and smaller compression compared to the higher frequency one because of its large ``mode inertia'' (a mode's tendency to remain unchanged, in analogue to the inertia of normal matter).
  On the other hand, there are two obvious features for $\delta\bar{P}_{\rm nad}$. First, the magnitude of (non-adiabatic) pressure fluctuations decreases with increasing frequency, which is in line with what shown in Fig.~\ref{fig:PSdeltaP}. Second, for both frequencies in Fig.~\ref{fig:Pandxi}, pressure fluctuations diminish drastically in the photospheric layers, where energy balance is controlled primarily by radiative processes and convective turbulence is comparatively small.
  
  Putting the two aspects together allows us to investigate the value of the work integral (also referred to as ``$PdV$ work'', where ``$P$'' stands for the pressure fluctuation and ``$dV$'' is the compression of fluid caused by oscillations) throughout the simulation domain. For lower frequency oscillations, weak compression is coupled with strong pressure fluctuations around and below the photosphere, whereas for higher frequency oscillations, major contributions to the $PdV$ work come from a small region around the photosphere, below which the compression of fluid is small, above which pressure fluctuations are small. 
  
  In summary, energy transfer from convective motions to oscillations is carried out by non-adiabatic effects such as convective turbulence and entropy fluctuations. The amount and rate of energy injection into the modes is governed by two main aspects, the magnitude of pressure fluctuation and the strength of local compression to fluid. The former is strong at low frequencies, and decays with increasing frequency (Fig.~\ref{fig:PSdeltaP}) while the latter exhibits an opposite trend (Fig.~\ref{fig:Pandxi}). As a result, excitation rates at low and high frequencies are limited by oscillations (local compression) and convection (pressure fluctuations), respectively. Finally, we note that the numerical results and main conclusions in this section are qualitatively in agreement with SN01 in general, although our solar model differs from theirs.

\section{Damping rates} \label{sec:damping}

\begin{figure*}
\subfigure[]{
\begin{overpic}[width=0.47\textwidth]{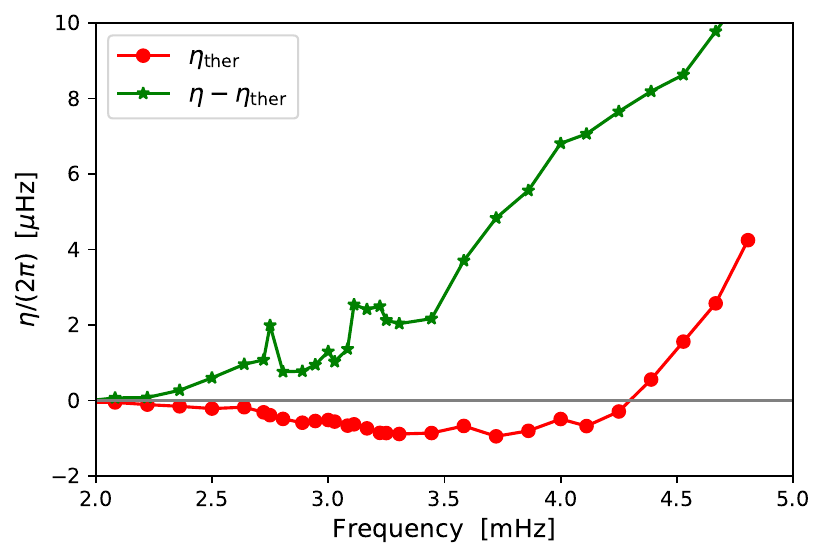}
\end{overpic}
\label{fig:eta_ther}
}
\subfigure[]{
\begin{overpic}[width=0.47\textwidth]{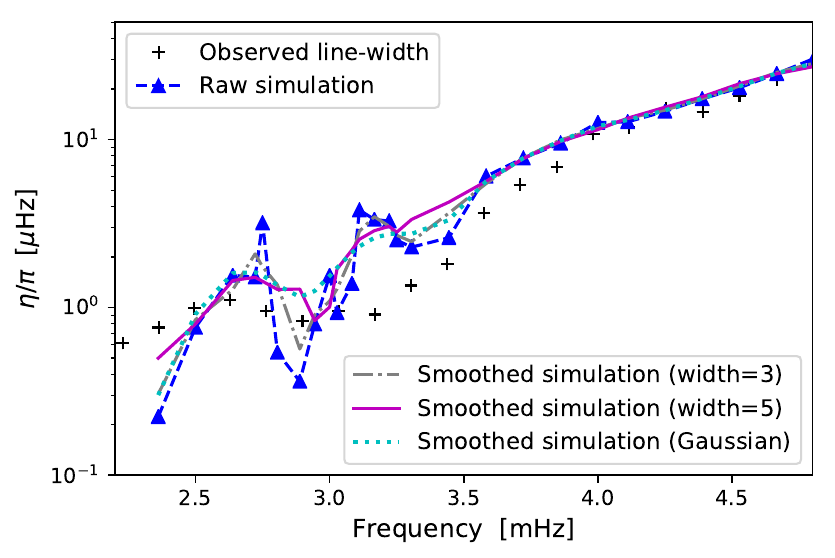}
\end{overpic}
\label{fig:eta}
}
\caption{
Linear damping rates at different cyclic frequencies for the Sun, computed from 3D simulation as described in Sect.~\ref{sec:damping}. Left panel: red dots are contribution to damping rate from thermal pressure fluctuation while total damping rate minus thermal pressure contribution (green asterisks) mainly reflect damping due to turbulent convection. Right panel: total damping rates from simulation (raw data in blue triangles, smoothed data in magenta, grey dash-dot and cyan dotted line) are divided by $\pi$ in order to compare with observed line widths (plus mark) from BiSON $l=0$ data \citep{1998MNRAS.298L...7C}. The magenta and grey curves are obtained by taking the running mean of the raw simulation data with a width of five and three data points respectively, whereas the cyan dotted line results from smoothing the raw simulation data by a Gaussian kernel with full width at half maximum equals to 0.25 mHz.}
\label{fig:damping}
\end{figure*}

  The stochastic excitation mechanism discussed in Sect.~\ref{sec:excitation} is responsible for the driving of p-mode oscillations. However, the amplitude of excited mode cannot grow infinitely large because it is limited by another effect called mode damping, the dissipation of mode kinetic energy to surroundings. Therefore, the final equilibrium oscillation amplitude results from the balance between stochastic excitation (energy gain) and mode damping (energy loss). The energy dissipation process is quantified by the \textit{damping rate $\eta$} which describes how fast is the $e$-folding (decay by a factor of $e$) of mode energy. In this section, we outline how damping rates at different frequencies are computed from first principles, and present our results for radial oscillations of the Sun.
  
  The (linear) damping rate can be derived from first order perturbation theory, with non-adiabatic effects as small perturbation (cf.~\citealt{2010aste.book.....A} Sect.~3.7),
\begin{equation} \label{eq:eta_num}
\eta = \frac{\omega \int_{y_{\rm bot}}^{y_{\rm top}} 
\mathfrak{Im}\left\{ (\delta\bar{\rho}^{*} / \bar{\rho}_0) \delta \bar{P}_{\rm nad} \right\} dy }
{4 m_{\rm mode} \vert \tilde{v}_y(R_{\rm phot}) \vert^2},
\end{equation}
where star symbol represents complex conjugate. $m_{\rm mode}$ and $\tilde{v}_y$ are mode mass per unit area and vertical velocity amplitude respectively, they are connected to mode kinetic energy (per unit surface area) by $m_{\rm mode} \vert \tilde{v}_y(R_{\rm phot}) \vert^2 = E_0$ (NS01 Eq.~63). The integral at numerator is the work integral which depend on (non-adiabatic) pressure and density fluctuation as well as the phase difference between them. The density fluctuation $\delta\rho$ quantifies the extent of local compression, it is related to mode displacement vector by the perturbed fluid continuity equation:
\begin{equation} \label{eq:conti}
\nabla \cdot \delta\vec{r} = -\frac{\delta\rho}{\rho_0}.
\end{equation}
Furthermore, the sign of $\eta$ is a criterion of mode stability: positive $\eta$ implies stable mode whose amplitude decays exponentially with time if no energy is supply from, for instance, convection; negative $\eta$, also called growth rate in this scenario, suggests that mode with finite amplitude will continue to drain energy from surrounding until the amplification is halted by nonlinear effects. Solar radial modes studied in this work are believed to be stable (\citealt{2015LRSP...12....8H} Sect.~6.3).

  It is worth noting that while the mode displacement vector (or equivalently, the eigenfunction) and density fluctuation is related through \eqref{eq:conti}, the \textit{adiabatic} eigenfunction $\xi_r$ obtained from \textsc{adipls} cannot be applied to the calculation of \textit{non-adiabatic} pressure fluctuation. To this end, all components in Eq.~\eqref{eq:eta_num}, except for $m_{\rm mode}$, must be extracted directly from simulation. Nevertheless, as first proposed by \cite{1998IAUS..185..199N}, it is challenging to separate coherent fluid motion from the turbulent convective processes in simulations. More specifically, ideally, density fluctuations should only contain the contribution from collective fluid motions (i.e.~oscillations). But in reality, owing to the complexity of physical processes occurring in the simulation domain, $\delta\bar{\rho}/\bar{\rho}_0$ computed from simulation consists not only pulsation signals but also the signature of ``convective noises''. This effect is particularly evident for modes that do not naturally emerge in simulations (i.e.~radial modes other than the three simulation modes shown in Fig.~\ref{fig:PSvy}). In order to obtain a $\delta\bar{\rho} / \bar{\rho}_0$ that cleanly reflects the contribution from mode eigenfunction, in other words, a coherent density fluctuation, we conduct numerical experiments that artificially drive radial mode at a particular frequency to large amplitude. The target mode will be prominent in the simulation box and distinguishable from ``convective noise''. 
   
   The artificial driving is achieved by modifying the boundary condition of the simulation. Namely, we impose a small time-dependent perturbation to thermal (gas plus radiation) pressure at the bottom boundary while keep the entropy constant (in first order) at the same time to ensure no extra energy is injected to the system. The applied thermal pressure perturbation varies sinusoidally with time and remains uniform over horizontal plane, since radial modes are the focus here.
   
   The perturbation at the bottom boundary will generate coherent fluid motion with the same frequency as the driving frequency, and amplify vertical velocity, density and pressure fluctuation to unrealistically large magnitudes. Nevertheless, we claim that such artificial driving would not compromise the reliability of our damping rate result because the rates of $\delta\bar{\rho}$, $\delta\bar{P}_{\rm nad}$ and $\tilde{v}_y$ enhancement are similar to each other, so that the artificial effect from ``mode driving'' largely cancels out between $(\delta\bar{\rho}^{*} / \bar{\rho}_0) \delta \bar{P}_{\rm nad}$ and $\vert \tilde{v}_y(R_{\rm phot}) \vert^2$ when we compute damping rate at the driving frequency using Eq.~\eqref{eq:eta_num}.
   
  Such numerical experiment is repeated at different driving frequencies in order to obtain theoretical damping rates as a function of cyclic frequency. In Fig.~\ref{fig:eta_ther} we separate the contribution to $\eta$ due to thermal pressure fluctuation,
\begin{equation}
\delta\bar{P}_{\rm ther}(t) = 
\bar{P}_{\rm ther,L}(t) - \bar{P}_{\rm ther,0,L}.
\end{equation}
$\delta\bar{P}_{\rm ther}$ stems from the divergence of radiative and convective fluxes (\citealt{1991LNP...388..195S} Sect.~3) which is most prominent near photosphere where the transition between radiative and convective heat transport occurs. For frequencies span from $\sim 2$ mHz to $\sim 4$ mHz, thermal pressure fluctuation is responsible for destabilizing modes, qualitatively in line with what found in \cite{1999A&A...351..582H}. On the other hand, another major contribution to mode damping is convective turbulence. The positive-definite $\eta - \eta_{\rm ther}$ indicate that turbulence tends to dissipate mode energy at all frequencies. That is to say, solar radial modes excited by turbulent convection is actually damped by the same effect, in accordance with the assertion in \cite{2015LRSP...12....8H} Sect.~6.3.

  Also, in Fig.~\ref{fig:eta}, our numerical results are compared with line width\footnote{When observing time is much greater than the mode $e$-folding time, damping rate is related to line width $\Gamma$ by $\Gamma = \eta / \pi$ \citep{2005MNRAS.360..859C}.} of $l=0$ modes collected by BiSON \citep{1998MNRAS.298L...7C}. Good agreements are found at high and intermediate frequencies. The observed damping rate plateau around 2.8 mHz is also well predicted. However, as can be seen from Fig.~\ref{fig:eta}, the accuracy of our results at low-frequency ($\nu \lesssim 2.5$ mHz) is restricted by the depth of simulation domain. As shown in Fig.~\ref{fig:xi}, low-frequency radial modes have considerable oscillation amplitude in solar interior. Because work integral is truncated at the bottom of simulation box in practice, contributions from deeper layers are omitted hence the final damping rates are systematically lower than observed values at low frequencies.

\section{Estimate velocity amplitude and $\nu_{\max}$ from simulation} \label{sec:Vandnumax}

\begin{figure}
\begin{overpic}[width=0.95\columnwidth]{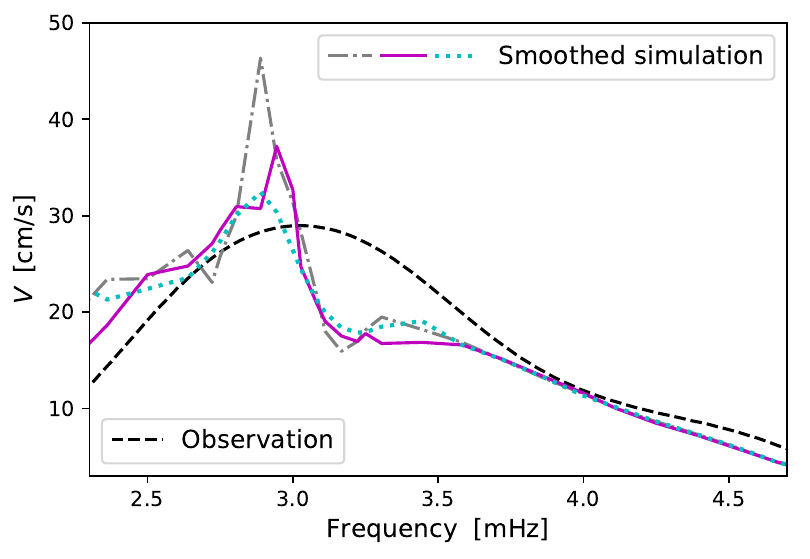}
\end{overpic}
\caption{Predicted photosphere velocity amplitude as a function of cyclic frequency, as evaluated using Eq.~\eqref{eq:Vamp}. Magenta, grey dash-dot and cyan dotted lines denote theoretical results from different smoothing options for damping rates (see Fig.~\ref{fig:eta}). The mean radial velocity amplitude derived from BiSON data \citep[black dashed line]{2008ApJ...682.1370K} has been smoothed and divided by the projection factor $0.724$ to enable a direct comparison between simulations and observation.}
\label{fig:V}
\end{figure}

  The observed oscillation amplitudes result from the balance between stochastic excitation and mode damping. With both of them available from the simulation, we are able to evaluate (theoretical) oscillation amplitude, then estimate the frequency of maximum oscillation power $\nu_{\max}$.
  
  For an observed mode in equilibrium state (in other word, constant amplitude), its energy gain and loss must be equal, that is
\begin{equation}
\mathcal{P}_{\rm exc} + \frac{dE_{\rm osc}}{dt} = 0.
\end{equation}
Here $\mathcal{P}_{\rm exc}$ is the excitation rate of this mode, and $E_{\rm osc}$ is its kinetic energy whose canonical form is (\citealt{2010aste.book.....A} Eq.~3.141)
\begin{equation} \label{eq:def_Eosc}
E_{\rm osc} = \frac{1}{2} M_{\rm mode} V_{\rm rms}^2,
\end{equation}
where $M_{\rm mode}$ is mode mass (not to be confused with mode mass per unit area $m_{\rm mode}$) and $V_{\rm rms}$ being root-mean-square (rms) velocity at photosphere. The evolution of kinetic energy for a damped oscillator follows $E_{\rm osc} \propto \exp(-2\eta t)$, therefore
\begin{equation} \label{eq:Eosc_eva}
E_{\rm osc} = \frac{\mathcal{P}_{\rm exc}}{2\eta}.
\end{equation}
Combining Eqs.~\eqref{eq:def_Eosc} and \eqref{eq:Eosc_eva} gives the expression of the rms velocity:
\begin{equation}
V_{\rm rms} = \sqrt{ \frac{\mathcal{P}_{\rm exc}}{M_{\rm mode} \eta} }.
\end{equation}
The mean kinematic velocity amplitude at the photosphere due to one oscillation mode is then
\begin{equation} \label{eq:Vamp}
V = \sqrt{2}V_{\rm rms} = \sqrt{ \frac{2\mathcal{P}_{\rm exc}}{M_{\rm mode} \eta} }.
\end{equation}

  Note that $V$ is not an asteroseismic observable hence cannot be compared with observations directly. What is obtained from analysing the Doppler shift of spectral lines (spectroscopic measurement of stellar oscillation) is the radial velocity. The source of this radial velocity is indeed the kinematic velocity of the fluid, whereas the quantity one measures is affected also by limb darkening and other geometric effects \citep{2010aste.book.....A}. These two kinds of velocities are connected by projection factor $S_{nlm}$ (also named spatial response function) that accounts for these effects (\citealt{1982MNRAS.198..141C} Eq.~4.1), 
\begin{equation} \label{eq:RV-Vamp}
\mathfrak{v}_{nlm} = S_{nlm} V_{nlm},
\end{equation}
where $\mathfrak{v}_{nlm}$ is the observed mean radial velocity amplitude of p-mode with quantum number $(n,l,m)$. \cite{1989MNRAS.239..977C} has shown that the projection factor for $l=0$ modes of the Sun is 0.724. Equations \eqref{eq:Vamp} and \eqref{eq:RV-Vamp} therefore enable comparison between the predicted velocity amplitude and the observed mean radial velocity amplitude.

  The theoretical $V$ is computed based on Eq.~\eqref{eq:Vamp}, with smoothed excitation and damping rates evaluated in Sect.~\ref{sec:er_result} and \ref{sec:damping}, respectively. Smoothed instead of raw simulation data are applied to calculate $V$ in order to avoid strong random fluctuation in the latter and to make $V$ comparable with observations. The power spectrum of the observed radial velocity is also smoothed. Mode mass is calculated from 1D patched solar model (see Sect.~\ref{sec:er_num}) using \textsc{adipls}. The thus computed velocity amplitude is shown in Fig.~\ref{fig:V}, together with the observed radial velocity spectrum taken from \cite{2008ApJ...682.1370K}.

  Moderate agreement between simulation and observation is found: velocity amplitude predicted from simulation is in the same order of magnitude as the observationally inferred value, and the overall shape of $V - \nu$ curve also resembles observation. In the mean time, we do notice that there are mismatches between the two, especially within $3 \lesssim \nu \lesssim 3.5$ mHz. In this frequency interval we underestimated excitation rates (Fig.~\ref{fig:er}) and overestimated damping rates (Fig.~\ref{fig:damping}). Errors on both aspects overlay then propagate into $V$ results. 

  In spite of the discrepancy between simulation and observation in the absolute magnitude of the velocity amplitude, we are still able to provide an estimate for $\nu_{\max}$ from the simulation. In the context of spectroscopic measurement of stellar oscillation, the frequency of maximum power is the corresponding frequency where the mean observed radial velocity reaches its maximum. This criterion is adopted in \cite{2008ApJ...682.1370K} where they obtain $\nu_{\max}=3.1$ mHz for the Sun. As an analogy, the maximum of theoretical $V$ then predicts a theoretical solar $\nu_{\max}$ which, from Fig.~\ref{fig:V}, is $\sim 3.0$ mHz. We note that the exact theoretical $\nu_{\max}$ value from the 3D solar simulation is somewhat ambiguous because it depends on how exactly the smoothing is performed (Fig.~\ref{fig:V}).
  
  Finally, we clarify why photosphere velocity should be computed via Eq.~\eqref{eq:Vamp}. On the surface it seems photosphere velocities are directly available from 3D simulation (Fig.~\ref{fig:PSvy}), then why calculate it semi-analytically? Here we emphasize neither photosphere velocity adopted from ``normal'' 3D simulation nor from artificial mode driving experiment is comparable with $V$. The reason for the latter is obvious -- artificial driving will amplify oscillation at a particular frequency to unrealistically large amplitude, therefore the absolute value of photosphere velocity in such numerical experiment has no physical meaning. (Damping rate value, however, is reliable because of the cancellation between  $(\delta\bar{\rho}^{*} / \bar{\rho}_0) \delta \bar{P}_{\rm nad}$ and $\vert \tilde{v}_y(R_{\rm phot}) \vert^2$, as discussed in Sect.~\ref{sec:damping}.)

  Using photosphere velocity directly from a ``normal'' 3D simulation is also inappropriate. From Fig.~\ref{fig:PSvy} it is clear that vertical velocity of the intermediate-frequency simulation mode is on the order of 1 km/s, much larger than the observed velocity amplitude of individual p-mode which is on the order of 1 m/s (Fig.~\ref{fig:V}). The mismatch between simulated and observed velocity results from the limited volume of the simulation domain. As demonstrated in previous sections, 3D solar simulation is able to predict realistic mode excitation and damping rates with their absolute value similar to what deduced from helioseismic observations. Because mode kinetic energy $E_{\rm osc}$ is dictated by excitation and damping processes (Eq.~\eqref{eq:Eosc_eva}), $E_{\rm osc}$ from the simulation is comparable to the actual kinetic energy of solar radial modes. However, in simulations, oscillations are confined in the simulation domain whereas for the Sun, radial modes oscillate throughout the entire solar surface and interior. Given the similarity in kinetic energy and the difference in cavity volume, it is apparent that the finite size of simulation will lead to larger oscillation amplitude. The velocity directly from the simulation is not a realistic representation of solar mode amplitude unless proper scaling is performed, as also discussed in \cite{2019A&A...625A..20B}.

\section{Conclusions}

  In this paper, we investigated how radial oscillations in the Sun are excited and damped based on 3D hydrodynamical simulation of solar near-surface region. Our simulation provides a realistic model of fluid motions and heat transport around photosphere. Its \textit{ab initio} nature also allows us to compute mode excitation and damping rates in an essentially parameter-free manner. 
  
  For the excitation rate, we adopted the theoretical framework developed by NS01 and SN01. Ingredients in the expression of excitation rate are calculated directly from simulation. Our numerical results demonstrate that mode excitation is weak at low frequencies, it increases rapidly with frequency and reaches its maximum between 2.5 mHz and 4 mHz, then start to decline at higher frequencies. It is also verified that mode excitation stems from the coupling between convection-induced pressure fluctuation and pulsation-induced fluid compression. Excitation rates computed in the current work is consistent with previous theoretical investigation (e.g., SN01) and corresponding solar observation.
  
  A novel numerical technique is applied to extract (linear) damping rates from simulations. In order to separate the fluctuation caused by pulsation from convective effects, we artificially drive a target radial oscillation to large amplitude, then compute damping rate from such simulation using analytical formula \eqref{eq:eta_num}. Broad agreement is achieved between simulation and observation, especially for higher frequency modes. The damping rate ``plateau'' around 2.8 mHz is also observed. What is more, by analysing different aspects that contribute to mode damping, we found thermal processes tend to destabilize radial modes with frequency between 2 mHz and 4 mHz while convective turbulence, which is responsible for mode driving, is also the main effect that dissipates them. This conclusion is in agreement with the findings by \citet{1999A&A...351..582H}, although the latter calculated mode damping in a radically different way.
  
  With both mode excitation and damping rates extracted from the model, it is possible to produce a prediction for the theoretical velocity spectrum, from which $\nu_{\max}$ can be estimated. The velocity amplitude is obtained by assuming exact balance between energy gain from stochastic excitation and energy loss by linear damping. Based on the synthetic velocity amplitude, we report, to our knowledge, the first pure theoretical $\nu_{\max}$ estimation for the Sun. Theoretical velocity amplitude and $\nu_{\max}$ are compared with observationally inferred values from \cite{2008ApJ...682.1370K} with an encouraging agreement.
  
  The major highlight of the current work is that all results are based on first principles. Given the 3D simulation of solar convection introduced in Sect.~\ref{sec:model}, the formulation we present does not depend on any free parameter that need to be calibrated from observation or other theoretical models. In addition, our method enables a detailed analysis of the interaction between convection and pulsations. From the simulation it is also possible to isolate from each other the different factors contributing to mode excitation or damping.
  In short, 3D numerical simulations provide full information about physical processes happened in solar near-surface region, some of them, such as the work integral, are difficult to probe by observation or traditional 1D models.
  
  On the other side, results from observation can be used to assess how well p-mode oscillations are modelled by 3D simulations, given that these two methods are completely independent. As mentioned in earlier sections, excitation and damping rates, as well as velocity amplitude evaluated from 3D simulation agree with corresponding observation in general, which indicates solar oscillations are overall properly simulated in this work. However, we caution that discrepancies between numerical results and solar observations do exist in both excitation and damping rates. The differences then propagate into synthetic velocity amplitude and theoretical $\nu_{\max}$. 
  These mismatches are indicative of shortcomings in our numerical simulations (or analytical formulation). Among them, the most notable one is the spacial size and time span of simulation. The finite depth in 3D model actually truncates the work integral, limiting it in the simulation domain. Thus additional contributions from outside the simulation box are neglected. Temporally, although 3D simulation used in this work is extremely long (compared to 3D solar model generated for other purposes such as spectra synthesis, which normally cover approximately only one hour of solar time \citep{2013A&A...557A..26M}) in time, the time sequence is still far from enough to resolve all radial modes excited in the Sun. Modelling of excitation and damping of modes would benefit from having a model that is more extended in depth and has a longer duration. Nevertheless, the main restriction in this respect remains computational time.
  Apart from the size of simulation, we also note that our analysis is strictly restricted to radial modes. In reality, the frequency of maximum oscillation power is determined from the full solar velocity spectrum which contains also non-radial p-modes. Hence, extracting $\nu_{\max}$ from radial modes only is a simplified approach. Nonetheless, we claim that this simplification might not be a significant flaw, because solar oscillation spectra are dominated by p-modes with degree $l=0-3$ (\citealt{2010aste.book.....A} Sect.~7.1.3) which are all radial or near-radial oscillations. Our formulation therefore should also hold approximately for these low-degree p-modes.

\acknowledgments

  The authors are grateful to Dennis Stello and Yaguang Li for reading and commenting on this manuscript. We thank also Luca Casagrande, Christoph Federrath, Mike Ireland, \AA ke Nordlund and Tim Bedding for valuable comments and fruitful discussions. YZ thanks the hospitality of Stellar Astrophysics Centre at Aarhus University during his visit. MA gratefully acknowledges funding from the Australian Research Council (grant DP150100250). Funding for the Stellar Astrophysics Centre is provided by The Danish National Research Foundation (Grant agreement no.:~DNRF106). This research was undertaken with the assistance of resources provided at the NCI National Facility systems at the Australian National University through the National Computational Merit Allocation Scheme supported by the Australian Government.

\end{document}